\documentclass[aps,prb,twocolumn,superscriptaddress,showpacs]{revtex4}
\usepackage{graphicx}
\begin{document}

\title{Retention of Two--Band Superconductivity in Highly Carbon--Doped MgB$_{2}$}

\author{H.\ Schmidt}
\affiliation{Materials Science Division, Argonne National
Laboratory, Argonne, IL 60439} \affiliation{Physics Division,
Illinois Institute of Technology, Chicago, IL 60616}

\author{K.E.\ Gray}
\affiliation{Materials Science Division, Argonne National
Laboratory, Argonne, IL 60439}

\author{D.G.\ Hinks}
\affiliation{Materials Science Division, Argonne National
Laboratory, Argonne, IL 60439}

\author{J.F.\ Zasadzinski}
\affiliation{Materials Science Division, Argonne National
Laboratory, Argonne, IL 60439} \affiliation{Physics Division,
Illinois Institute of Technology, Chicago, IL 60616}

\author{M.\ Avdeev}
\affiliation{Materials Science Division, Argonne National
Laboratory, Argonne, IL 60439}

\author{J.D.\ Jorgensen}
\affiliation{Materials Science Division, Argonne National
Laboratory, Argonne, IL 60439}

\author{J.C.\ Burley}
\affiliation{Materials Science Division, Argonne National
Laboratory, Argonne, IL 60439}

\date{\today}

\begin{abstract}
Tunneling data on MgB$_{1.8}$C$_{0.2}$ show a reduction in the
energy gap of the $\pi$--bands by a factor of two from undoped
MgB$_{2}$ that is consistent with the $T_{c}$ reduction, but
inconsistent with the expectations of the dirty limit.
Dirty--limit theory for undoped MgB$_2$ predicts a single gap
about three times larger than measured and a reduced $T_c$
comparable to that measured. Our heavily--doped samples exhibit a
uniform dispersion of C suggestive of significantly enhanced
scattering, and we conclude that the retention of two--band
superconductivity in these samples is caused by a selective
suppression of \emph{interband} scattering.
\end{abstract}


\maketitle

The simple binary compound, MgB$_2$,\cite{jNagamatsu01} appears to
be an elegant example of a two--band superconductor with a high
$T_c$ of 39~K, in which superconductivity due to strong
electron--phonon coupling in the two--dimensional (2D)
$\sigma$--bands is induced into the 3D $\pi$--bands by a
combination of interband electron--phonon coupling and weak
interband quasiparticle scattering. Several key signatures of this
include the closing of two energy gaps
\cite{hSchmidt02a,hSchmidt03,pSzabo01,fGiubileo01,mIavarone02,rsGonnelli02b}
at the same $T_c$, anomalous features in the specific heat and its
field dependence \cite{fBouquet01a} and the direct observation of
interband scattering in the tunneling spectrum.\cite{hSchmidt02a}

However, two distinct superconductive gaps should only be seen in
the clean limit, while in Ref.\ \cite{ayLiu01} a single
intermediate gap with a lower $T_c$ is predicted in the dirty
limit. The observed weak dependence of $T_c$ on residual
resistivity is not obviously consistent with this latter
expectation.\cite{iiMazin02b} Mazin \emph{et al.}
\cite{iiMazin02b} have used supercell band--structure calculations
to conclude that the two gaps can be preserved in the presence of
enhanced scattering provided the \emph{interband} impurity
scattering is extremely weak. As will be shown below, introducing
additional scattering centers by doping enables us to reduce the
mean free path below the coherence length and thereby offers a
direct experimental test of the dirty--limit prediction.

In a two--band superconductor the term \emph{dirty limit} needs to
be used carefully, though. The mean free path is reduced by both
\emph{intra}-- and \emph{interband} scattering, but only
\emph{interband} scattering is expected to homogenize the two gaps
and result in effective one--band behavior.\cite{iiMazin02b} Thus
there are two different scenarios for doped MgB$_2$: (i)
impurities strongly enhance \emph{interband} scattering, in which
case the small gap, $\Delta_\pi$, is expected to increase until it
merges with the large gap, $\Delta_\sigma$, at an intermediate
value, or (ii) \emph{interband} impurity scattering is minimally
affected and two distinct gaps are preserved. In scenario (ii),
the ratio $2\Delta_\pi\over k_BT_c$ for the $\pi$--band is
expected to remain small ($\sim1.5$ in undoped\cite{hSchmidt02a}
MgB$_2$), whereas it would increase in scenario (i). Thus
determination of $2\Delta_\pi\over k_BT_c$ alone allows us to
decide between these scenarios, and in this paper we present
tunneling data showing that both $\Delta_\pi$ (the smaller gap)
and $T_c$ are simultaneously reduced by about a factor of two upon
$\sim10$\% C--doping. This represents direct experimental evidence
for the selective suppression of \emph{interband} impurity
scattering relative to \emph{intraband} scattering.

Several groups
\cite{wMickelson02,raRibeiro03,tTakenobu01,aBharathi02a,zhCheng02,mParanthaman01a,iMaurin02,kPapagelis02}
have investigated the doping of MgB$_2$ with C with conflicting
results concerning the C solubility limit, the composition
dependence of $T_c$ and the possible presence of phase separation.
Much of the discrepancy is possibly a synthesis problem due to the
slow kinetics of C incorporation. Mickelson \emph{et
al.},\cite{wMickelson02} and more recently Ribeiro \emph{et
al.},\cite{raRibeiro03} have been able to overcome the diffusion
problem by using a C containing starting material, B$_4$C, and
obtained samples with sharp diffraction peaks and superconducting
transitions.  The synthesis using B$_4$C yields multiphase samples
containing another borocarbide, MgB$_2$C$_2$, since the solubility
limit of C in MgB$_2$ is less than the 20~atom\% present in the
starting B$_4$C.

\begin{figure}
\vskip0.05in
\includegraphics[scale=0.57]{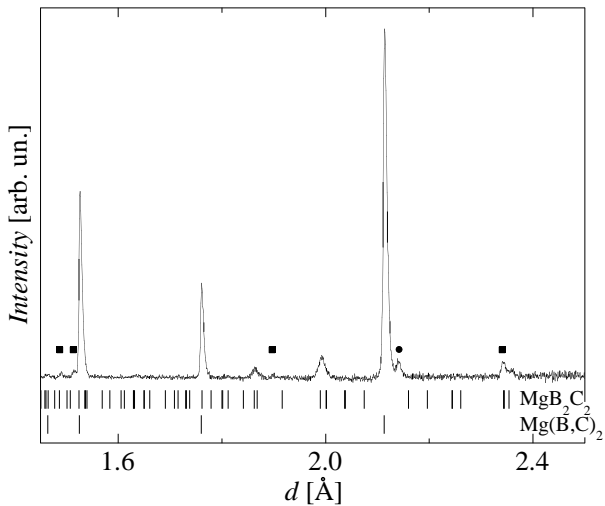}
\vskip-0.16in \caption{\label{fig1} Partial neutron powder
diffraction pattern for the MgB$_{1.8}$C$_{0.2}$ sample.  The
solid squares and circle show the diffraction lines for Cd and V,
respectively, that are instrumental in nature.  The only other
extraneous diffraction lines are from MgB$_2$C$_2$, present at
about .07 mole fraction.}
\end{figure}

\begin{figure}
\vskip0.05in
\includegraphics[scale=0.565]{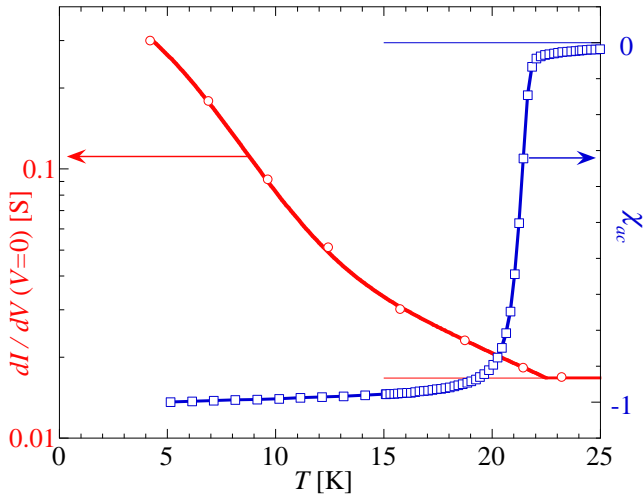}
\vskip-0.16in \caption{\label{fig1b}(color online) Determination
of the bulk--$T_c$ from normalized ac susceptibility (squares),
and the junction--$T_c$ from the zero--bias conductance due to
Josephson coupling (circles). Excellent agreement precludes
reduced surface superconductivity.}
\end{figure}

Our samples were synthesized from high--purity Mg, graphite, and
Eagle--Picher $^{11}$B enriched to 99.52\%.  A low carbon content
boron carbide sample with a B$_{0.85}$C$_{0.15}$ composition was
first made by arc melting the B and graphite together.  This
material was ground to a coarse powder, mixed with excess Mg and
fired at 1000$^{\circ}$C for 5~hr as described in Ref.
\cite{dgHinks02}. After the excess Mg was removed by heating to
400$^{\circ}$C in vacuum, powder neutron diffraction revealed only
sharp diffraction peaks (Fig.\ \ref{fig1}) and the ac
susceptibility showed a sharp transition with a midpoint $T_c$ of
21.5~K (Fig.\ \ref{fig1b}). Note the absence of broadening of the
diffraction lines precludes phase separation that was reported in
Ref. \cite{iMaurin02,kPapagelis02} for highly C--doped MgB$_2$.
Rietveld analysis of the neutron diffraction data yield lattice
constants $a=3.0528$~{\AA} and $c=3.5219$~{\AA} with a .07 mole
fraction of MgB$_2$C$_2$ impurity. The $a$--axis is much reduced
from 3.0849~{\AA} for the pure material, while the $c$--axis is
relatively unchanged. MgB$_2$C$_2$ is still present in our samples
(although at a lower concentration then in Refs.\
\cite{wMickelson02,raRibeiro03}) since our starting C content is
still larger than the solubility limit of C in MgB$_2$. From the
initial carbon content of the boron carbide sample and the final
MgB$_2$C$_2$ impurity content, the carbon content of the C--doped
MgB$_2$ sample is determined to be MgB$_{1.8}$C$_{0.2}$, i.e., a
10~atom\% substitution of C for B, with an accuracy of
$\pm2$~atom\%. This defines the solubility limit that agrees with
recent results of Avdeev \emph{et al.}\cite{mAvdeev03}

\begin{figure}
\vskip-0.055in
\includegraphics[scale=0.57]{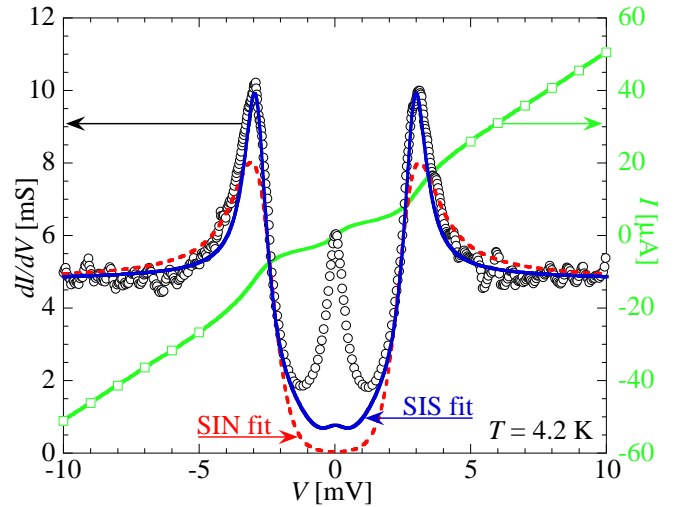}
\vskip-0.1in \caption{\label{fig2}(color online) $I(V)$
characteristic (squares) for a relatively low resistance junction
in C--doped MgB$_2$ taken at 4.2~K. Also shown are the numerical
derivative (circles) and calculated fits for single electron
tunneling in both the SIS (solid line) and
superconductor--insulator--normal metal (SIN) (dashed line)
configurations. The appearance of the Josephson pair--tunneling
peak at zero voltage and the generally excellent agreement with
the solid line establish the SIS configuration.}
\end{figure}

Our tunneling data are on
superconductor--insulator--superconductor (SIS) break junctions
\cite{hSchmidt02a,hSchmidt03} and they commonly identify the
smaller gap in the 3D $\pi$--bands of C--doped MgB$_2$. An example
is shown in Fig.\ \ref{fig2} where the current--voltage
characteristic, $I(V)$, is plotted as the line with open squares
along with its differential conductance, $dI\over dV$, that was
generated numerically and is plotted as open circles. Our
tunneling apparatus \cite{lOzyuzer98} uses Au tips that readily
attach to a piece of MgB$_2$ and afterwards the single tunnel
junction formed between this attached piece of MgB$_2$ and the
bulk sample predominates the spectra, i.e., has by far the highest
resistance.\cite{hSchmidt02a,hSchmidt03} The data exhibit rather
sharp coherence peaks near $\pm3$~mV that are characteristic for
SIS junctions,\footnote{Note the qualitative disagreement with an
SIN fit using a gap of 2.6~meV and virtually zero lifetime
broadening, $\Gamma$. At 4.2~K, thermal effects alone would
broaden the peaks to about half their observed height, as shown by
the dashed line in Fig.\ \ref{fig2}.} and an overall spectral
shape that is well--reproduced by an SIS fit (solid line in Fig.\
\ref{fig2}) using $\Delta_\pi=1.45$~meV and $\Gamma=0.25$~meV.
This ratio of $\Gamma/\Delta_\pi\sim20$\% is consistent with our
tunneling data on undoped MgB$_2$.\cite{hSchmidt02a,hSchmidt03}

\begin{figure}
\vskip0.1in
\includegraphics[scale=0.57]{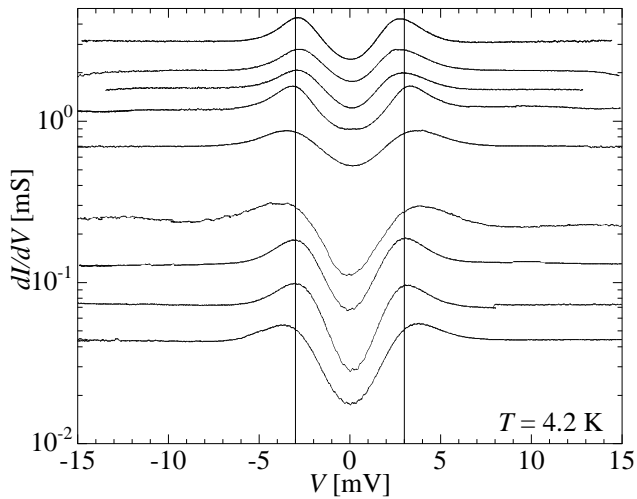}
\caption{\label{fig3} A compendium of SIS data taken on two
different samples, \emph{on a logarithmic scale}, indicating the
reproducibility of the location of the coherence peak (energy gap)
and the general shape of the data. These higher--resistance
junctions do not show the Josephson pair tunneling at zero bias
seen in Fig. \ref{fig2}.}
\end{figure}

The Josephson pair--tunneling peak at zero bias is the most
prominent difference with the SIS calculation that only considers
single--electron, quasiparticle tunneling.\footnote{Note that the
SIS calculation reveals a small hump at zero bias due to tunneling
of thermally--activated quasiparticles. This is seen in
high--resistance, low--$\Gamma$ tunnel junctions in undoped
MgB$_2$,\cite{hSchmidt02a,hSchmidt03} but like the Josephson
current it is a feature that is only found in SIS junctions.  The
observed zero--bias peak seen in Fig.\ \ref{fig2} is too large to
be explained by thermal activation, but agrees with the Josephson
pair tunneling that is seen in such low--resistance
($\sim210$~$\Omega$ in this case) junctions in undoped MgB$_2$.}
Due to the exceptional strength of this feature in
lower--resistance tunnel junctions, and its intimate relation to
superconductivity, the disappearance of this zero--bias peak at
higher temperature is a very accurate test of the junction--$T_c$.
We traced this feature to high temperatures in a junction that at
4.2~K showed almost 20 times the normal--state background
conductance at zero bias, and find it to merge into a flat
background at $T_c\sim22-23$~K (Fig.\ \ref{fig1b}).\footnote{Note
that the zero--bias conductance signal is \emph{not proportional}
to the Josehson current, $I_J$, itself and its temperature
dependence is not expected to follow $I_J(T)$. However, the
disappearance of both quantities coincides and allows for an
unambiguous determination of $T_c$.} This is in excellent
agreement with the bulk--$T_c$ from magnetization that for
comparison is also shown in the same figure, and allows us to draw
two conclusions. First, the zero--bias peak is indeed due to
Josephson currents, thereby providing solid corroboration of SIS
tunneling. Second, the junction--$T_c$ is equivalent to the bulk
value, confirming that the tunneling results are not affected by
surface effects. Higher--resistance junctions do not show the
Josephson pair tunneling at zero bias, as is expected. A selection
of these, taken on two different samples, are shown in Fig.\
\ref{fig3} to indicate the reproducibility of the location of the
coherence peak (energy gap) and the general shape of the data.

To explain why we see only the smaller energy gap in the 3D
$\pi$--band, we follow the reasoning of Ref.\ \cite{hSchmidt02a}.
Because of the exponential decay of the electron wavefunction in a
barrier, tunneling currents are generally dominated by electron
momenta that are within a small angle (typically $\sim15$ degrees)
of the junction normal. Getting within 15$^\circ$ of the
$ab$--plane in a randomly formed junction is highly improbable;
thus while all tunneling data on MgB$_2$ observe the smaller gap
in the 3D $\pi$--bands, the 2D $\sigma$--band is seen only
occasionally. When present, contributions from the 2D
$\sigma$--band are easily distinguished in SIS data
\cite{hSchmidt03} but are not seen in the data of Figs.\
\ref{fig2} and \ref{fig3}, consistent with more than 95\% of the
junctions we have observed in undoped MgB$_2$. We thus have an
unambiguous determination of the smaller gap in the 3D
$\pi$--bands.

There is a disparity between this small gap value,
$\Delta_\pi\sim1.5$~meV, and a large critical temperature,
$T_c\sim22$~K, that is crucial experimental evidence for two--band
superconductivity. The gap ratio, ${2\Delta_\pi\over k_BT_c}\sim
1.53$, is severely reduced from the weak--coupling limit of
$\sim3.53$, a disparity that would only be increased by
strong--coupling effects. This can only be explained by the
presence of a second, larger energy gap, i.e., two--band behavior.
We note, that $2\Delta_\pi\over k_BT_c$ for C--doped MgB$_2$,
$\sim1.53$, is roughly the same as in undoped MgB$_2$, $\sim1.49$,
i.e.\ critical temperature and gap scale together with C--doping.

We therefore conclude that two--band superconductivity persists in
C--doped MgB$_2$, consistent with specific heat data that show an
anomaly reminiscent of that attributed to two--band effects in
undoped MgB$_2$.\cite{raRibeiro03} We further conclude that this
result gives direct experimental evidence for the absence of
\emph{interband} impurity scattering, i.e. scenario (ii). Although
a comparable reduction to $T_c=25.4$~K was predicted for the dirty
limit,\cite{aBrinkman02} i.e., strong \emph{interband} scattering,
this would be connected to a single intermediate gap,
$\Delta\sim4.1$~meV,\cite{aBrinkman02} which is about three times
the observed value. From the qualitatively different behavior of
the small gap we conclude that the lowering of $T_c$ in
MgB$_{1.8}$C$_{0.2}$ is caused by changes in the electronic
structure rather than by enhanced scattering.

This reduction of $T_c$ may be understood by the model of
covalently bonded hexagonal B nets which are ionically bonded by
Mg$^{2+}$ ions.\cite{jmAn01,jKortus01} The attraction of the
$\pi$--electrons to the Mg$^{2+}$ ions lowers the energy of the
$\pi$-- relative to the $\sigma$--bands, transferring electrons
from the $\sigma$-- to the $\pi$--bands. It is these resultant
holes in the $\sigma$--bands that are responsible for the high
$T_c$. Substituting C into MgB$_2$ might be expected to change the
effective carrier concentration both by direct charge doping and
by indirect charge transfer $\pi\rightarrow\sigma$ through a
modification of the ionic system. To first order the large
decrease in the $a$--axis on C doping results from direct electron
doping of the bonding $\sigma$--bands, causing a net decrease in
the hole content. The unchanged $c$--axis indicates little effect
on the ionic bonding between Mg$^{2+}$ ions and the B nets. Thus
it is probably the lowered density of states (DOS) in the
$\sigma$--bands resulting from direct charge doping by C that
causes the large $T_c$ decrease. Medvedeva \emph{et al.}
\cite{niMedvedeva01b} have calculated the DOS for several
different dopants in MgB$_2$. For C they find a reduced DOS and
lower $T_c$, predicting further that $T_c$ should fall to zero at
a doping level of about 0.085 electrons/B atom as the
$\sigma$--band falls below the Fermi energy. Yan \emph{et al.}
\cite{yYan02b} also find a reduced $T_c$ and a smaller B--B
interatomic distance on C doping. All these predictions are in
qualitative agreement with our data, although we find that
superconductivity persists at 22~K for $\sim0.1$ electrons/B atom.

The retention of two--band superconductivity at such a high doping
level is surprising, as it requires the absence of sizable
\emph{interband} impurity scattering. Again, there are essentially
two scenarios that allow us to understand this behavior: (a)
doping does not increase \emph{any} scattering, or (b) doping
increases \emph{intraband} scattering only, while \emph{interband}
scattering is unaffected. In scenario (a) the lack of any
scattering implies the C forms a superlattice. Scenario (b)
requires the selective suppression of \emph{interband} scattering
relative to \emph{intraband} scattering, and below we will present
evidence that supports scenario (b).

For impurities as dilute as the 10\% C for $^{11}$B in our
samples, the possibility of local ordering, e.g., staged
replacement of, on average, every 5th B layer with an ordered
(BC)$_{0.5}$ layer cannot be ruled out, but the long--range order
needed for scenario (a) would not be expected. Testing for this is
difficult since the contrast between C and $^{11}$B is weak by any
scattering technique. Electron scattering offers the highest
sensitivity to ordering because the diffraction occurs from tiny
single crystals in the powder sample. Transmission electron
diffraction was performed on crystallites aligned along the [001]
and [100] zone axes, and no evidence of any ordering of dopant
ions was found. We therefore conclude, that C enters the lattice
as randomly distributed impurities, and that the sharp $T_c$ and
the narrow powder diffraction peaks (see Fig.\ \ref{fig1}) are the
result of a solubility limit for C in MgB$_2$ (that in the
presence of excess C would be expected to result in a homogenous
doping just at this limit) rather than a well--ordered C
superlattice.

In the absence of C--ordering, the dopant atoms are expected to
act as scattering centers, and we may estimate the effective mean
free path from the C--concentration. Using a $2\times2\times1$
MgB$_2$ supercell containing one C atom, equivalent to
$1/8=12.5$\% impurities, the average C--C in--plane distance is
$\sim0.6$~nm, about one order of magnitude less than the
anisotropic coherence lengths in undoped MgB$_2$ (that are 4 and
10~nm \cite{lLyard02}). Such a significantly reduced mean free
path is borne out by the increase in ${dH_{c2}\over dT}(T_c)$ by
about a factor of two upon similar C--doping.\cite{raRibeiro03}

We therefore conclude that the retention of two--band
superconductivity in the presence of heavy C--doping is a result
of the selective suppression of \emph{interband} impurity
scattering, scenario (b). (Note that the absence of sizable
\emph{interband} scattering is a non--trivial prerequisite for the
observation of two--band superconductivity even in undoped
MgB$_2$.) Mazin \emph{et al.} \cite{iiMazin02b} studied scattering
in MgB$_2$ theoretically and argued that \emph{interband} impurity
scattering would be suppressed due to the disparity between
$\sigma$-- and $\pi$--wavefunctions. Applied to the case of
C--doped MgB$_2$, this suggests that the two--band nature of
superconductivity (that requires the absence of sizable
\emph{interband} scattering) can be retained, while at the same
time other properties (that depend on \emph{intraband} scattering)
may be suggestive of the dirty limit, e.g., show increased
resistivity and ${dH_{c2}\over dT}(T_c)$.

In conclusion, our tunneling results on MgB$_{1.8}$C$_{0.2}$
demonstrate the retention of two--band superconductivity in the
presence of significant scattering, consistent with specific heat
measurements of Ribeiro \emph{et al.}\cite{raRibeiro03} We explain
this by the selective suppression of \emph{interband} scattering,
as suggested by Mazin \emph{et al.}\cite{iiMazin02b} Our
experiment shows that this suppression is a surprisingly robust
feature of MgB$_2$ and strong enough to retain the two--band
character even in the presence of a C--doping content sufficient
to reduce $T_c$ by a factor of two.

The authors acknowledge discussions with D.J.\ Miller, and
technical support from S.\ Short and H.\ Zheng. This research is
supported by the U.S.\ Department of Energy, Basic Energy
Sciences---Materials Sciences, under contract \#
W--31--109--ENG--38.

\end{document}